# Extending the photon energy coverage of an x-ray self-seeding FEL via the reverse taper enhanced harmonic generation technique

Kaiqing Zhang, Zheng Qi, Chao Feng*, Haixiao Deng, Dong Wang*, and Zhentang Zhao

*Shanghai Institute of Applied Physics, Chinese Academy of Sciences, Shanghai 201800, China*

\* Corresponding authors.

E-mail addresses: fengchao@sinap.ac.cn (C. Feng), wangdong@sinap.ac.cn (D. Wang).

**Abstract:** In this paper, a simple method is proposed to extend the photon energy range of a soft x-ray self-seeding free-electron laser (FEL). A normal monochromator is first applied to purify the FEL spectrum and provide a coherent seeding signal. This coherent signal then interacts with the electron beam in the following reverse tapered undulator section to generate strong coherent microbunchings while maintain the good quality of the electron beam. After that, the pre-bunched electron beam is sent into the third undulator section which resonates at a target high harmonic of the seed to amplify the coherent radiation at shorter wavelength. Three dimensional simulations have been performed and the results demonstrate that the photon energy gap between 1.5 keV and 4.5 keV of the self-seeding scheme can be fully covered and 100 GW-level peak power can be achieved by using the proposed technique.

## 1. Introduction

The successful operations of free-electron lasers (FELs) have opened up a new chapter to the exploration of chemistry, biology, material science as well as many other scientific frontiers due to its remarkable characteristics of high brightness, short pulse and completely transverse coherence. Self-amplified spontaneous emission (SASE) [1-3] is currently the main operation mode of x-ray FELs and has been proved reliable and successful to provide high peak power, ultra-short light pulses with good spatial coherence. Nevertheless, SASE FEL has relative wide bandwidth and poor longitudinal coherence because the signal initiating the radiation starts from the electron shot noise [4]. It can be barely used in the scientific experiments requiring temporal coherence. At this circumstance, several seeded FEL schemes with external seed lasers have been developed to improve the longitudinal coherence. The simplest way is directly seeding a FEL with a high-harmonic generation (HHG) source [5]. While the HHG direct-seeding technique has been successfully demonstrated in the VUV wavelengths [6], it can hardly generate x-ray radiation due to the limitation of the HHG power at this wavelength range. An alternative way for seeding FEL at shorter wavelength is adopting high harmonic generation schemes such as high-gain harmonic generation (HGHG) [7, 8] or echo-enable harmonic generation [9-11]. Unfortunately, the limitation of the energy modulation amplitude prevents the possibility of reaching very short wavelength in a single stage setup. More concerns also come from the noise amplification through the harmonic multiplication process, which may spoil heavily the generated x-ray radiation properties. It is therefore hardly to push the output wavelength to the sub-nanometer region by using external seeding methods.

Instead of seeding with external lasers, the self-seeding technique [12-16] implements a monochromator together with a bypass chicane to obtain a monochromatic light from SASE itself and then amplify it to saturation. The undulator is divided into two parts by the monochromator. The first undulator section is used to generate a normal SASE radiation pulse that is interrupted well before saturation. Then the monochromator is employed to purify the spectrum and provide a coherent seeding signal at short wavelength, while the bypass chicane is used to delay the electron beam and wash out the microbunchings formed in the first undulator section. After that the seeding signal and the electron beam are simultaneously sent into the second undulator section to interact with each other and produce an intense coherent x-ray pulse.

Self-seeding FEL can be distinguished as soft x-ray self-seeding and hard x-ray self-seeding depending on the chosen of monochromator materials for different photon energy ranges. Generally, the grating-based monochromator [13] can be applied for the photon energy range of 0.7keV to 1.5keV, while diamond-based [14] monochromator is suitable for the photon energy range of 4.5keV to 10keV. Both the soft x-ray and hard x-ray self-seeding have been demonstrated at SLAC in recent years [15, 16]. To date, self-seeding is serving as one of most reliable configuration to provide high peak power, ultra-short x-ray light pulses with extraordinary coherence both transversely and longitudinally. However, there still exists a photon energy gap between 1.5keV-4.5keV that cannot be covered by present self-seeding schemes due to the lack of suitable materials for the monochromator. Previous studies showed a proper scheme to cover this photon energy gap by cascading the self-seeding with the HGHG [17]. However, in this scheme, a relative complex setup with separated seed amplifier and modulator is required to mitigate the beam quality degradation in the long modulator.

In this paper, we propose a novel method that combines the reverse undulator taper and harmonic generation techniques to extend the photon energy coverage of a self-seeding FEL. The proposed scheme utilizes a baseline configuration of a self-seeding FEL and does not require the installation of any additional hardwares in the undulator system. The proposed technique can be easily implemented at already existing or planned x-ray FEL facilities to generate x-ray radiation pulses with consecutively tuning wavelength from soft x-ray to hard x-ray region. This paper is organized as follows: the principles of the reverse taper enhanced harmonic generation technique are introduced in Sec. 2. Using a typical soft x-ray self-seeding parameters, three-dimensional simulation results are presented in Sec. 3. And finally some conclusions comments are given in Sec. 4.

## 2. Principles

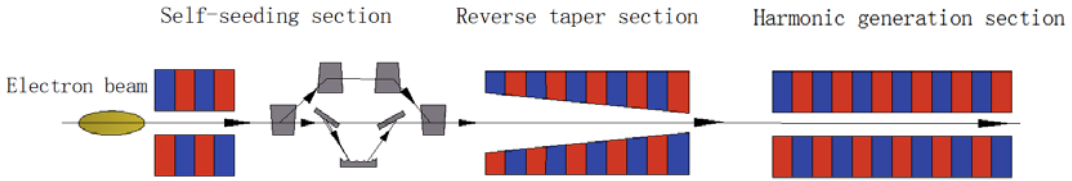

Fig. 1. Schematic layout of the reverse taper enhanced harmonic generation scheme.

The schematic layout of the proposed technique is shown in Fig.1. The undulator system consists of three undulator sections and a bypass chicane. The electron beam from the linac first passages through a short undulator in the self-seeding section to generate a SASE pulse with low output intensity. This radiation pulse is then sent through the grating monochromator to select a monochromatic light which works as the seed laser for the following radiation processes. The bypass chicane is used to adjust the approaching trajectory of the electron beam and compensate the radiation time delay induced by the monochromator. Meanwhile the michrobunching formed in the former SASE undulator can be smeared out by the chicane. After that the electron beam and the selected monochromatic light pulse are sent into the second undulator with reverse taper, which can imprint strong coherent microbunchings on the electron beam without significantly degrading the beam quality. Eventually, the pre-bunched electron beam radiates full power in the third undulator section (harmonic generation section), tuned to the resonance of a target high harmonic of the seed.

The reverse undulator taper technique has been initially proposed for obtaining high degree of circular polarization in x-ray FELs [18], which has been experimental demonstrated at the LCLS recently [19]. Different from a normal undulator taper technique, the magnetic field of the reversed tapered undulator is increased along the undulator segments, producing fully microbunched electron beam at the fundamental wavelength while efficiently suppress the powerful linearly polarized background. We will show below that that the reverse taper technique together with a coherent seed can be used to generate coherent microbunhcing that contains amount of Fourier components at high harmonics of the seed.

For a FEL in the high gain linear regime, the bunching factor $b$ and the complex amplitude of the harmonic of the energy modulation $b_p$ as a function of the normalized FEL power $\hat{\eta}$ can be simply calculated by (see the Appendix):

$$|b|^2 \simeq |\hat{C}|^2 \hat{\eta}, \quad |b_p|^2 \simeq |\hat{C}|\hat{\eta}, \quad (1)$$

where $\hat{\eta} = P/\rho P_{beam}$, $P$ is the FEL power at the undulator length $z$, $\rho$ is the Pierce parameter [20, 21], $P_{beam}$ is the electron beam power and $\hat{C}$ is the normalized detuning parameter. While Eq. (1) has exactly the same form as Eq. (A13) in Ref. [18], $\hat{\eta}$ is quite different due the initial monochromatic seed, which holds the ability to suppress the shot noise and helps to generate coherent microbunching. For a linear tapered undulator, the normalized detuning parameter is proportional to the taper strength $\beta$:

$$\hat{C} = \beta \hat{z} \quad (2)$$

and

$$\beta = -\frac{\lambda_w}{4\pi\rho^2}\frac{K(0)}{1+K(0)^2}\frac{dK}{d\hat{z}}, \quad (3)$$

where $\lambda_w$ is the undulator period, $\hat{z} = 4\pi\rho z/\lambda_w$, $K$ is the undulator parameter and $K0$ is the initial value of $K$ at the entrance of the undulator. Under the condition of a reverse tapered undulator with large negative detuning parameter, $\hat{C} < 0$, $|\hat{C}| \gg 1$, the bunching factor will change slightly with $\beta$ according to the Ref. [18], but the normalized radiation pulse energy will be suppressed significantly. According to Eq. (1), the bunching factor is proportional to $\hat{C}$ and the energy modulation amplitude is proportional to $\sqrt{\hat{C}}$. For a given value of bunching factor, the radiation induced energy spread growth in the reverse tapered undulator will be much smaller than that in a normal undulator. Thus the pre-bunched electron beam can be used again in the following undulator section for the harmonic generation.

To show the possible photon energy coverage of the proposed technique, here we adopt the practical parameters of a soft x-ray facility, as shown in Table. 1, to carry out some theoretical estimations. A 5 GeV electron beam with initial slice energy spread of 500 keV, peak current of 3 kA and normalized slice emittance of 0.6 mm-mrad is adopted to drive a soft x-ray self-seeding FEL with tunable central photon energy between 0.7keV and 1.5keV. In order to cover the photon energy range of 1.5-4.5 keV, we only need to tune the resonant wavelength of the third undulator section of the proposed scheme to the 2[nd] (1.4-3keV) and 3[rd] (2.1-4.5keV) harmonics of the seed. Based on the Xie's model [21], the calculated saturation power and photon energy coverage for different harmonics are shown in Fig. 2.

One can find that the output power for the proposed scheme is at GW-level over the photon energy range of 0.7-4.5 keV.

Table 1 Main parameters of a soft x-ray FEL facility

| Parameter | Value |
|---|---|
| Electron beam energy | 5 GeV |
| Energy spread | 0.01% |
| Peak current | 3 kA |
| Bunch length (rms) | 100 fs |
| Normalized emittance | 0.6 mm-mrad |
| Mono. central energy | 1.5 keV |
| Mono. resolution power | 10000 |
| Mono. power efficiency | 0.038 |
| FEL photon energy | 3-4.5 keV |
| Undulator period | 23.5 mm |
| SASE undulator length | 20 m |
| Reverse tapered undulator length | 20 m |
| Radiate undulator length | 80 m |

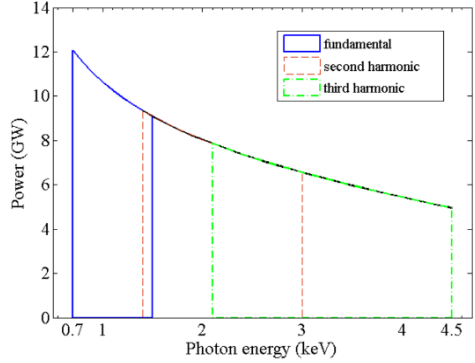

Fig. 2. The calculated saturation power and wavelength coverage of the fundamental, second harmonic and third harmonic radiation of the proposed scheme.

## 3. Three-dimensional simulations

In order to illustrate the optimization method and possible performances of the proposed technique, three-dimensional simulations were performed with GENESIS [22] based on the parameters present in table. 1. The undulator period is chosen to be 23.5 mm, the total undulator length is about 65 m. The self-seeding section is operated for the generation of a monochromatic light at 1.5 keV. An ideal Gaussian profile electron beam without considering the effect of the microbunching instabilities [23] is used in our simulation. In the first undulator section (20 m), the FEL is operated in the exponential gain regime to avoid the growth of the energy spread. Meanwhile the x-ray power after the monochromator should be at least two order of magnitude larger than the shot noise power in the downstream undulator. The design of the monochromator is similar with that in Ref. [24]. The resolution of the monochromator is chosen to be $10^{-4}$ to select a single spike in the spectrum. And about 3.8% x-ray power can be preserved due to the diffraction and reflection processes in the monochromator. The spectrum and the temporal distribution of the radiation pulses before and after the monochromator are shown in Fig. 3. One can find in Fig. 3 that the normalized spectrum bandwidth is reduced from $5 \times 10^{-3}$ to about $3 \times 10^{-5}$ by the monochromator. The peak power is reduced by about three orders of magnitude, from 250 MW to about 0.18 MW. Both the spectrum and temporal profiles become rather smooth after the monochromator.

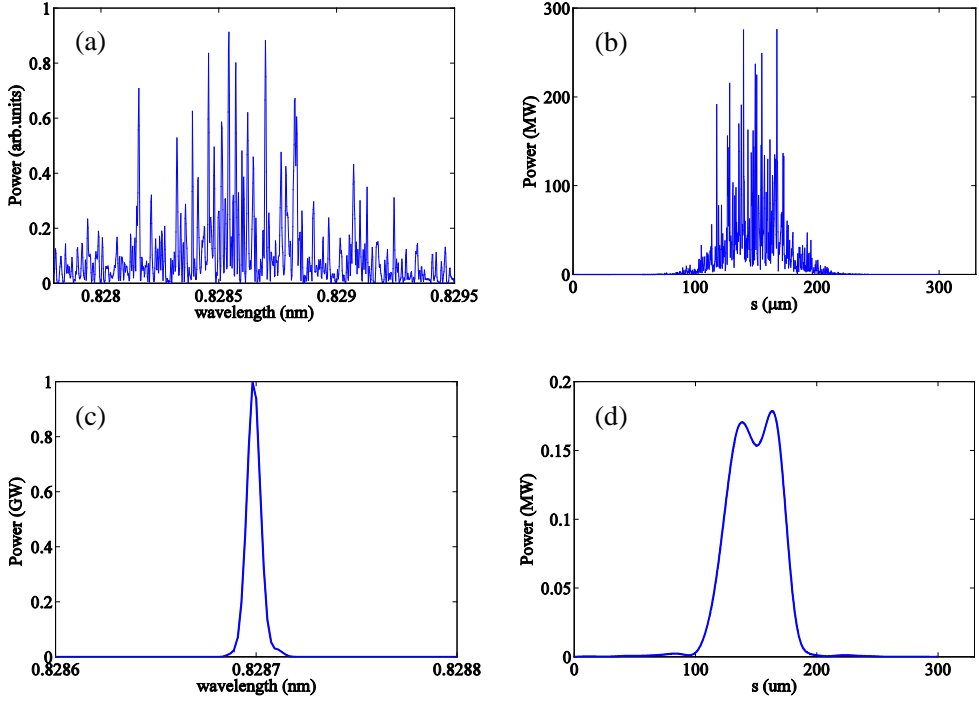

Fig. 3. The spectrum and temporal distributions of the radiation pulse before (top plots) and after (bottom plots) the monochromator.

The monochromatic seed and the refreshed electron beam are then directly sent in to the subsequent reverse tapered undulator. As mentioned above, the reverse tapered undulator is introduced in our scheme to obtain a well bunched electron beam with suppression growth of the beam energy spread. However, the saturation length in the reversed tapered undulator will increase correspondingly if the taper strength $|\beta|$ is too large. We need to optimize the taper strength to ensure that the saturation power is sufficiently suppressed while the saturation length does not increase obviously. Fig. 4 shows the simulation results of the saturation power and the relative saturation length variations as a function of $\beta$ in the second undulator section. It is clearly seen that the saturation power declines significantly when $\beta$ is smaller than -0.4, while the gain length will increase quickly when the reverse taper strength is smaller than -0.6. Thus we can reasonably conclude that the taper strength is appropriate to be chosen between -0.6 and -0.4 to achieve a suitable saturation length and a suppressed saturation power. In the following simulations, the taper strength is chosen to be -0.4.

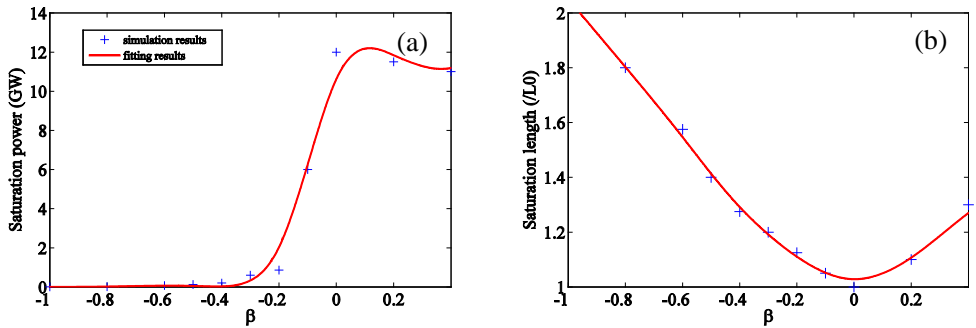

Fig. 4. Simulation results for (a) the saturation power and (b) relative saturation length as a function of $\beta$. L0 is the saturation length for $\beta = 0$.

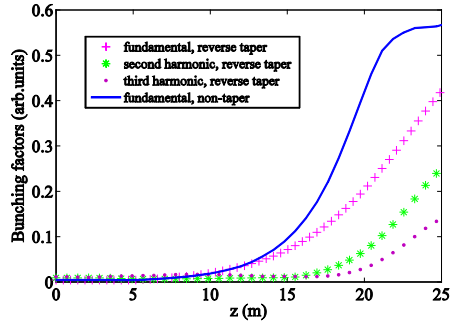

Fig. 5. Bunching factors evolutions along the second undulator section.

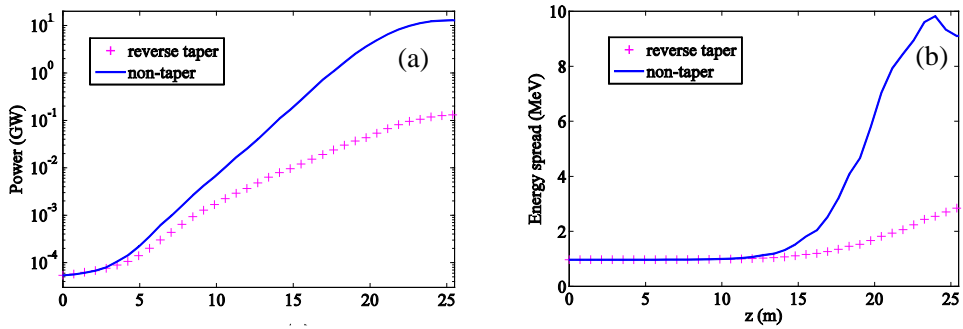

Fig. 6. Simulation results for (a) the radiation peak power and (b) the electron beam energy spread evolutions along the second undulator section.

The simulation results of the bunching factors, radiation peak power and electron beam energy spread evolutions along the reverse tapered undulator are shown in Fig. 5 and Fig. 6. For comparison purpose, simulations for the normal undulator case have also been performed. These simulation results indicate that the fundamental bunching factor at the exit of the reverse tapered undulator is almost the same as the non-tapered one, while the saturation power is suppressed by about two orders of magnitude, from about 10 GW to about 100 MW. The radiation induced energy spread has been reduced from about 9.2 MeV to about 2.6 MeV. Besides the fundamental bunching factor, the electron beam also contains abundant high harmonic components. As shown in Fig. 5, the second harmonic bunching factor is increased to about 0.28 and the third harmonic bunching factor is increased to about 0.14 at the exit of the reverse tapered undulator. These bunching factors are sufficient large to initiate the coherent high harmonic radiation in the following undulator.

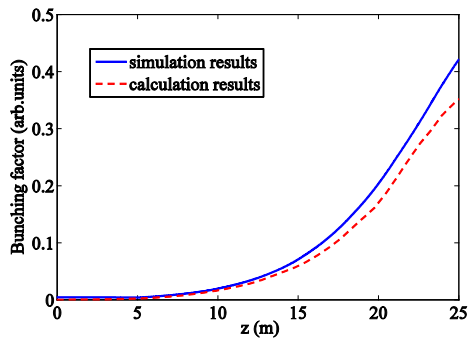

Fig. 7. Comparison of the simulation and calculation results for the fundamental bunching factor

evolutions along the reverse tapered undulator.

The evolution of the bunching factor along the reverse tapered undulator can also be theoretically calculated by using Eq. (1). Fig. 7 gives the calculation results with the simulation parameters. One can find that the calculation curve fit quite well with the three-dimensional simulation result.

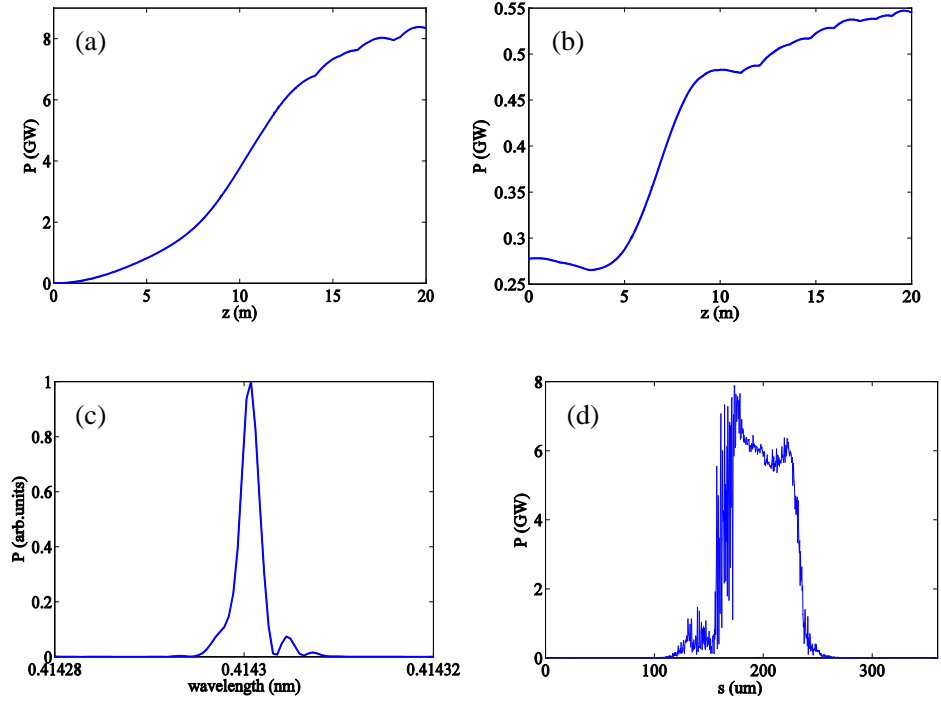

Fig. 8. Simulation results for the second harmonic generation: radiation peak power (a) and bunching factor (b) as a function of the undulator axis, radiation pulse (c) and corresponding spectrum (d) at saturation.

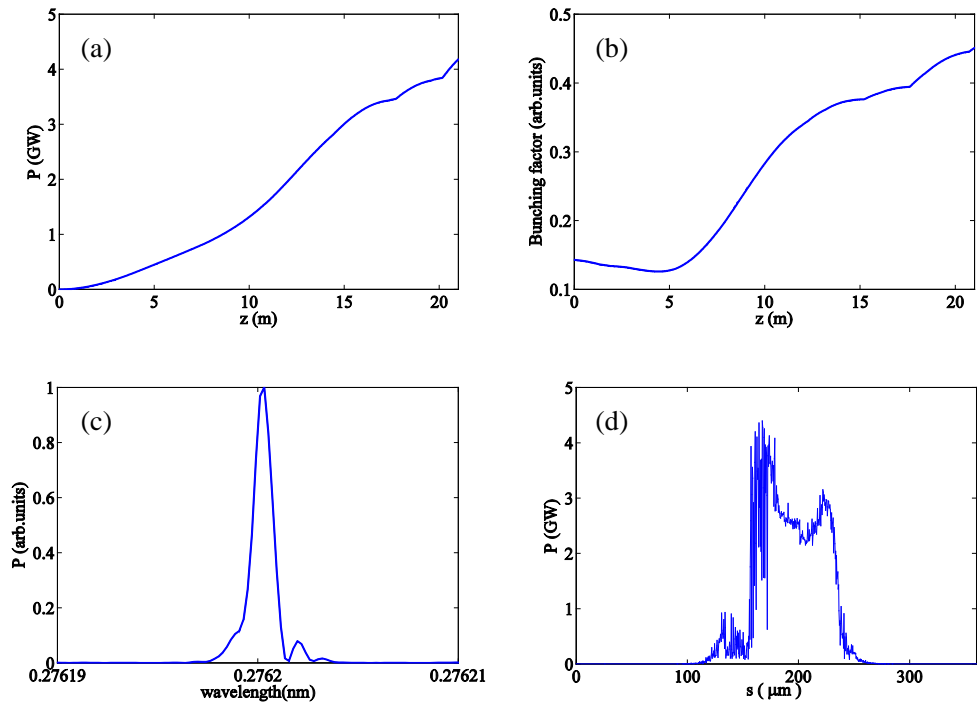

Fig. 9. Simulation results for the third harmonic generation: radiation peak power (a) and bunching factor (b) as a function of the undulator axis, radiation pulse (c) and corresponding spectrum (d) at saturation.

Finally, the pre-hunched electron beam is sent into the third undulator section which is resonant at a target high harmonic of the seed to lase. We carried out simulations for both the second and the third harmonic cases. The simulation results are illustrated in Fig. 8 and Fig. 9. The second harmonic radiation (3 keV) gets saturation at around 15 m of the undulator with a saturation power of about 7.5 GW, while the third harmonic radiation (4.5 keV) gets saturation at around 20 m with a saturation power of about 4 GW. The spectral bandwidths (FWHM) of the second and third harmonic radiation pulse are both quite close to the Fourier transform limit.

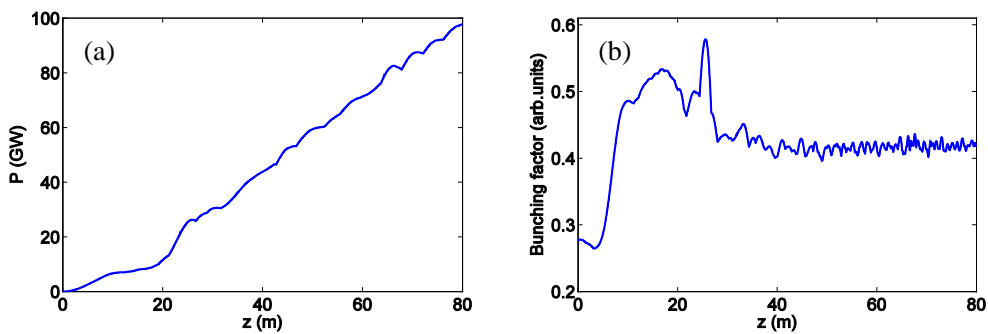

Fig. 10. Second harmonic radiation power and the bunching factor evolutions along the third undulator section with taper.

It is well known that the power extraction efficiency of the monochromatic radiation pulse can be significantly increased by using the normal taper technique. Fig. 10 shows the simulation result for the gain curve of second harmonic radiation in a tapered undulator. The length of the third undulator section is increased to 80 m. A taper strength of 0.2 is adopted after 15 m of the undulator. One can find from Fig. 10 that the peak power can reach 100 GW at the end of the undulator. The bunching factor of about 0.4 can be well maintained along the tapered undulator.

## 4. Conclusions and perspective

In summary, an easily-to-implement method has been proposed to extend the photon energy range of a soft x-ray self-seeding FEL. Theoretical analysis and numerical simulations are given and the results demonstrate that the reverse tapered undulator can provide modulated electron beam with strong microbunchings and small energy spread. This kind of electron beam can be directly sent into the following undulator for the generation of GW-level coherent radiation pulse at second and third harmonics of the seed, which can fully cover the photon energy gap between 1.5 keV-4.5 keV. The output peak power can be further enhanced to 100 GW by using the normal taper technique.

The proposed technique has a relatively simple configuration and can be easily implemented in the present soft x-ray self-seeding FEL facilities around the world, which will make the self-seeding scheme to be much more reliable and useful in generation of consecutive x-ray radiation pulses from soft x-ray to hard x-ray region. The proposed technique can also be applied to a hard x-ray self-seeding FEL to further extend the photon energy coverage, where no suitable monochromatic materials can reach.

It should be point out here that some practical limiting factors that might affect the performance of the proposed scheme, such as the effects from the microbunching instability in the electron beam, are not taken into account in this paper. Further investigations on these topics are ongoing.

## APPENDIX

The detail deduction of reverse taper theory without external seed has been done in the appendix of Ref. [18]. Following the notations of Ref. [18], here we repeat the derivation process considering the external coherent seed.

The evolution of the electronic field of the amplified electromagnetic wave can be derived as:

$$\tilde{E}''' + 2i\hat{C}E''' - \hat{C}\tilde{E}' = i\tilde{E} \qquad (A1)$$

where $\hat{C}$ is detuning parameter, the solution for the equation A1 can be expressed as:

$$\tilde{E} = \sum_{j=1}^{3} C_j \exp(\lambda_j \hat{Z}), \qquad (A2)$$

where $C_j$ are constant coefficents, $\lambda_j$ are the solutions of eigenvalue equation:

$$\lambda(\lambda + i\hat{C})^2 = i. \qquad (A3)$$

The evolution of the field amplitude and its second and third derivatives with respect to $\hat{z}$ can be calculates by:

$$\begin{bmatrix} E \\ E' \\ E'' \end{bmatrix}_{\hat{Z}} = M(\hat{z}|0) \begin{bmatrix} E \\ E' \\ E'' \end{bmatrix}_0 \qquad (A4)$$

The matrix elements of $M(\hat{z}|0)$ are:

$M_{11} = \lambda_2\lambda_3 B_1 + \lambda_1\lambda_3 B_2 + \lambda_1\lambda_2 B_3$,
$M_{12} = -(\lambda_2+\lambda_3)B_1 - (\lambda_1+\lambda_3)B_2 + (\lambda_1+\lambda_2)B_3$,
$M_{13} = B_1 + B_2 + B_3$,
$M_{21} = \lambda_1\lambda_2\lambda_3 M_{13}$,
$M_{22} = -\lambda_1(\lambda_2+\lambda_3)B_1 - \lambda_2(\lambda_1+\lambda_3)B_2 + \lambda_3(\lambda_1+\lambda_2)B_3$,
$M_{23} = \lambda_1 B_1 + \lambda_2 B_2 + \lambda_3 B_3$,
$M_{31} = \lambda_1\lambda_2\lambda_3 M_{23}$,
$M_{32} = -\lambda_1^2(\lambda_2+\lambda_3)B_1 - \lambda_2^2(\lambda_1+\lambda_3)B_2 + \lambda_3^2(\lambda_1+\lambda_2)B_3$,
$M_{33} = \lambda_1^2 B_1 + \lambda_2^2 B_2 + \lambda_3^2 B_3$, $\qquad (A5)$

where

$$B_1 = \frac{\exp(\lambda_1 \hat{z})}{(\lambda_1-\lambda_2)(\lambda_1-\lambda_3)},$$

$$B_2 = \frac{\exp(\lambda_2 \hat{z})}{(\lambda_2-\lambda_1)(\lambda_2-\lambda_3)}, \qquad (A6)$$

$$B_3 = \frac{\exp(\lambda_3 \hat{z})}{(\lambda_3-\lambda_1)(\lambda_3-\lambda_2)}.$$

The relation between the derivatives $\tilde{E}'$, $\tilde{E}''$ and bunching $b(\hat{z})$ and the complex amplitude of harmonic of energy modulation $b_p(\hat{z})$ can be written as:

$$\frac{\tilde{E}'}{E_0} = -2b, \quad \frac{\tilde{E}''}{E_0} = -2i(\hat{C}b - b_p), \qquad (A7)$$

where $E_0$ is the normalizing factor of the field amplitude.

Different from Ref [18], in the proposed scheme, a monochromatic seed exits at the entrance of the reverse tapered undulator. The initial conditions should be written as

$$\tilde{E}(0) = A, \frac{\tilde{E}'(0)}{E_0} = -2b(0), \frac{\tilde{E}''}{E_0} = -2i\hat{C}b(0) \qquad (A8)$$

where $A$ is the amplitude of the initial electric filed. Thus we can get the radiation field and its

derivatives at any position of the undulator according to Eq. (A3):

$$\frac{\tilde{E}(\hat{z})}{E_0} = \frac{M_{11}A}{E_0} - 2M_{12}(\hat{z}|0)b(0) + 2i\hat{C}M_{13}(\hat{z}|0)b(0),$$

$$\frac{\tilde{E}(\hat{z})'}{E_0} = \frac{M_{21}A}{E_0} - 2M_{22}(\hat{z}|0)b(0) + 2i\hat{C}M_{23}(\hat{z}|0)b(0), \qquad (A9)$$

$$\frac{\tilde{E}(\hat{z})'}{E_0} = \frac{M_{31}A}{E_0} - 2M_{32}(\hat{z}|0)b(0) + 2i\hat{C}M_{33}(\hat{z}|0)b(0).$$

In the high gain linear regain, Eq. (A9) can be simplified as:

$$\frac{\tilde{E}(\hat{z})}{E_0} = \frac{\lambda_2\lambda_3 B_1 A}{E_0} + 2B_1(\lambda_2+\lambda_3 + 2i\hat{C})b(0),$$

$$\frac{\tilde{E}(\hat{z})'}{E_0} = \lambda_1 \frac{\tilde{E}(\hat{z})}{E_0}, \qquad (A10)$$

$$\frac{\tilde{E}(\hat{z})''}{E_0} = \lambda_1^2 \frac{\tilde{E}(\hat{z})}{E_0},$$

Plugging Eq. (A7) into Eq. (A10), we arrive

$$\frac{\tilde{E}(\hat{z})}{E_0} = \frac{\lambda_2\lambda_3 B_1 A}{E_0} + 2B_1(\lambda_2+\lambda_3 + 2i\hat{C})b(0),$$

$$b(\hat{z}) = -\frac{1}{2}\lambda_1 \frac{\tilde{E}(\hat{z})}{E_0}, \qquad (A11)$$

$$b_p(\hat{z}) = \frac{i}{2}\lambda_1(\lambda_1 + i\hat{C})\frac{\tilde{E}(\hat{z})}{E_0}.$$

By using the definition of $\hat{\eta} = |E|^2/4E_0^2$, we finally get:

$$|b|^2 \simeq |\hat{C}|^2 \hat{\eta}, \quad |b_p|^2 \simeq |\hat{C}|\hat{\eta}. \qquad (A12)$$